\documentclass[preprintnumbers,amsmath,amssymbm,prd]{revtex4}
\usepackage{epsfig}
\usepackage{graphicx}
\usepackage{amssymb}

\begin{document}
\title{Reissner-Nordstr\"om black holes supporting non-minimally coupled massive scalar field configurations}
\author{Shahar Hod}
\affiliation{The Ruppin Academic Center, Emeq Hefer 40250, Israel}
\affiliation{ }
\affiliation{The Hadassah Institute, Jerusalem 91010, Israel}
\date{\today}

\begin{abstract}
\ \ \ It has recently been demonstrated that static spatially
regular scalar fields, which are non-minimally coupled to the
electromagnetic field of a charged central black hole, can be
supported in the exterior regions of the black-hole spacetime. In
the present paper we use {\it analytical} techniques in order to
study the physical and mathematical properties of the externally
supported linearized scalar field configurations (scalar 'clouds')
in the dimensionless large-mass regime $\mu r_+\gg1$ (here $\mu$ and
$r_+$ are respectively the proper mass of the supported scalar field
and the outer horizon radius of the central supporting black hole).
In particular, we derive a remarkably compact analytical formula for
the discrete resonant spectrum
$\{\alpha_n(\mu;Q/M)\}_{n=0}^{n=\infty}$ which characterizes the
dimensionless coupling parameter of the composed
black-hole-non-minimally-coupled-linearized-massive-scalar-field
configurations. The physical significance of this resonant spectrum
stems from the fact that, for a given value of the dimensionless
black-hole electric charge $Q/M$, the fundamental (smallest)
eigenvalue $\alpha_0(\mu)$ determines the critical existence-line of
the composed black-hole-massive-field system, a boundary line which
separates non-linearly coupled hairy
charged-black-hole-massive-scalar-field configurations from bald
Reissner-Nordstr\"om black holes. The analytical results derived in
this paper are confirmed by direct numerical computations.
\end{abstract}
\bigskip
\maketitle

\section{Introduction}

The canonical no-hair theorems presented in
\cite{Bek1,Her1,Hodstationary} have revealed the physically
interesting fact that, in the composed Einstein-Maxwell-scalar
theory, static scalar field configurations cannot be supported in
the exterior regions of asymptotically flat black-hole spacetimes
with spatially regular horizons. This physically intriguing property
also characterizes black-hole spacetimes in which the scalar fields
are non-minimally coupled to the Ricci curvature scalar of the
spacetime \cite{BekMay,Hod1}.

Intriguingly, it has recently been demonstrated in the physically
interesting works \cite{Herch,Herr} that spatially regular massless
scalar field configurations which are characterized by a non-minimal
coupling of the form $f(\phi)F_{\mu\nu}F^{\mu\nu}$ to the
electromagnetic Maxwell tensor [see Eq. (\ref{Eq4}) below] {\it can}
be supported in spherically symmetric asymptotically flat charged
black-hole spacetimes. This phenomenon, which is known by the name
black-hole spontaneous scalarization, has also been studied in
\cite{aaa1,aaa2} in the physically interesting context of massive
scalar field configurations which are non-minimally coupled to the
electromagnetic tensor of the charged black-hole spacetime.

As explicitly proved in \cite{Herch,Herr}, if the non-trivial
scalar-field-electromagnetic-field coupling function is
characterized by the weak-field functional behavior
$f(\phi)=1+\alpha\phi^2+O(\phi^4)$, then the bald (scalar-less)
charged Reissner-Nordstr\"om black-hole spacetime is a valid
solution of the field equations in the trivial $\phi\equiv0$ limit.
This is a physically desirable property of the non-trivially coupled
Einstein-Maxwell-scalar theory. Here $\alpha>0$ is a dimensionless
physical parameter which determines the strength of the non-minimal
coupling between the supported scalar field and the electromagnetic
field of the charged black-hole spacetime.

The numerical results presented in the physically interesting works
\cite{Herch,Herr,aaa1,aaa2} have demonstrated that, for a given
value of the dimensionless charge-to-mass ratio $Q/M$ \cite{Noteqq}
of the black-hole spacetime, the non-trivially coupled
Einstein-Maxwell-scalar system is characterized by the existence of
a critical {\it existence-line} $\alpha=\alpha(\mu;Q/M)$ which marks
the boundary between hairy
charged-black-hole-non-minimally-coupled-massive-scalar-field
configurations and bald (scalar-less) Reissner-Nordstr\"om
black-hole spacetimes (here $\mu$ is the proper mass of the
non-minimally coupled scalar field). In particular, the critical
existence-line of the composed system corresponds to spatially
regular linearized field configurations which are supported by a
central charged Reissner-Nordstr\"om black hole (the term `scalar
clouds' is usually used in the physics literature
\cite{Hodlit,Herlit} in order to describe these linearized scalar
field configurations which sit on the critical existence-line of the
system).

It is important to emphasize the fact that, as nicely demonstrated
in \cite{Herch,Herr}, the critical existence-line of the
black-hole-field system is universal in the sense that different
scalar-field-electromagnetic-field coupling functions $\{f(\phi)\}$
with the same weak-field behavior,
$f(\phi)=1+\alpha\phi^2+O(\phi^4)$, are characterized by the same
functional behavior $\alpha=\alpha(\mu;Q/M)$ of the critical
existence-line.

The main goal of the present paper is to study, using {\it
analytical} techniques, the physical and mathematical properties of
the composed Einstein-Maxwell-non-minimally-coupled-scalar-field
theory in the dimensionless regime $M\mu\gg1$ of large field masses.
In particular, we shall derive a remarkably compact analytical
formula for the critical existence-line $\alpha=\alpha(\mu;Q/M)$
which characterizes the
Reissner-Nordstr\"om-black-hole-linearized-massive-scalar-field
cloudy configurations. Interestingly, the analytically derived
resonance formula [see Eq. (\ref{Eq31}) below] for the composed
black-hole-field system would provide a simple {\it analytical}
explanation for the {\it numerically} discovered \cite{aaa1,aaa2}
monotonic functional behavior of the relation
$\alpha=\alpha(\mu;Q/M)$ along the critical existence-line of the
system.

\section{Description of the system}

We shall study the physical and mathematical properties of
linearized massive scalar field configurations (scalar clouds) which are
non-trivially coupled to the electromagnetic field of a charged
Reissner-Nordstr\"om black hole. The line element of the spherically
symmetric charged black-hole spacetime can be expressed in the form
\cite{Noteunits}
\begin{equation}\label{Eq1}
ds^2=-h(r)dt^2+{1\over{h(r)}}dr^2+r^2(d\theta^2+\sin^2\theta
d\phi^2)\  ,
\end{equation}
where
\begin{equation}\label{Eq2}
h(r)=1-{{2M}\over{r}}+{{Q^2}\over{r^2}}\  .
\end{equation}
Here $M$ and $Q$ are respectively the black-hole mass and its
electric charge. The black-hole horizon radii $\{r_+,r_-\}$ are
determined by the polynomial equation $h(r=r_{\pm})=0$, which yields
\begin{equation}\label{Eq3}
r_{\pm}=M+(M^2-Q^2)^{1/2}\  .
\end{equation}

The composed
charged-black-hole-non-minimally-coupled-massive-scalar-field system
is characterized by the action \cite{Herch,Herr,aaa1,aaa2}
\begin{equation}\label{Eq4}
S=\int
d^4x\sqrt{-g}\Big[R-2\nabla_{\alpha}\phi\nabla^{\alpha}\phi-2\mu^2\phi^2-f(\phi){\cal
I}\Big]\ ,
\end{equation}
where the non-trivial coupling between the scalar field $\phi$ and
the electromagnetic Maxwell tensor $F_{\mu\nu}$ of the central
charged black hole is induced by the source term
\begin{equation}\label{Eq5}
{\cal I}=F_{\mu\nu}F^{\mu\nu}\  .
\end{equation}
The coupling function $f(\phi)$ of the supported massive scalar
field configurations is characterized by the universal quadratic
behavior \cite{Herch,Herr,aaa1,aaa2}
\begin{equation}\label{Eq6}
f(\phi)=1+\alpha\phi^2\
\end{equation}
in the weak-field regime, where the dimensionless expansion constant
$\alpha$ is the physical coupling parameter of the composed black-hole-field
theory. We shall henceforth assume $\alpha>0$.

The action (\ref{Eq4}), when varied with respect to the wave
function of the massive scalar field, yields the differential
equation \cite{Herch,Herr,aaa1,aaa2}
\begin{equation}\label{Eq7}
\nabla^\nu\nabla_{\nu}\phi={1\over4}f_{,\phi}{\cal I}\  .
\end{equation}
Substituting into (\ref{Eq7}) the line element (\ref{Eq1}) of the
curved black-hole spacetime and using the field decomposition
\cite{Notelm}
\begin{equation}\label{Eq8}
\phi(r,\theta,\phi)=\sum_{lm}{{\psi_{lm}(r)}\over{r}}Y_{lm}(\theta)e^{im\phi}\
,
\end{equation}
one finds that the spatial behavior of the static non-minimally
coupled massive scalar field configurations, which are supported by
the central charged Reissner-Nordstr\"om black hole, is determined
by the ordinary differential equation
\begin{equation}\label{Eq9}
{{d^2\psi}\over{dy^2}}-V\psi=0\  ,
\end{equation}
where the tortoise coordinate $y$ in the Schr\"odinger-like equation
(\ref{Eq9}) is related to the radial coordinate $r$ by the compact
differential relation \cite{Notemap}
\begin{equation}\label{Eq10}
{{dr}\over{dy}}=h(r)\  .
\end{equation}
Here \cite{Herch,Herr,aaa1,aaa2}
\begin{equation}\label{Eq11}
V(r)=\Big(1-{{2M}\over{r}}+{{Q^2}\over{r^2}}\Big)\Big[\mu^2+{{l(l+1)}\over{r^2}}+{{2M}\over{r^3}}-{{2Q^2}\over{r^4}}
-{{\alpha Q^2}\over{r^4}}\Big]\
\end{equation}
is the effective potential of the composed
black-hole-non-minimally-coupled-massive-scalar-field system.

In the next section we shall use analytical techniques in order to
determine the discrete resonant spectrum
$\{\alpha_n(\mu,l,M,Q)\}_{n=0}^{n=\infty}$ of the dimensionless
physical parameter $\alpha$. This resonant spectrum is determined by
the Schr\"odinger-like radial differential equation (\ref{Eq9}) with
the following boundary conditions \cite{Herch,Herr,aaa1,aaa2}
\begin{equation}\label{Eq12}
\psi(r=r_+)<\infty\ \ \ \ ; \ \ \ \ \psi(r\to\infty)\to
{{1}\over{r}}e^{-\mu r}\ .
\end{equation}
These physically motivated boundary conditions at the outer
black-hole horizon and at spatial infinity correspond to spatially
regular bound-state massive scalar field configurations which are
supported by the central charged black hole.

\section{The discrete resonant spectrum of the composed
charged-black-hole-linearized-massive-scalar-field system: A WKB analysis}

In the present section we shall derive a remarkably compact
analytical formula for the discrete resonant spectrum
$\{\alpha_n(\mu,l,M,Q)\}_{n=0}^{n=\infty}$ which characterizes the
composed charged-black-hole-linearized-massive-scalar-field
configurations in the dimensionless large-mass regime
\begin{equation}\label{Eq13}
M\mu \gg \text{max}\{1,l\}\  .
\end{equation}

As we shall now show explicitly, the Schr\"odinger-like equation
(\ref{Eq9}), which determines the radial functional behavior of the
spatially bounded non-minimally coupled massive scalar field
configurations in the charged black-hole spacetime (\ref{Eq1}), is
amenable to a WKB analysis in the large-mass regime (\ref{Eq13}). In
particular, a standard second-order WKB analysis of the
Schr\"odinger-like radial equation (\ref{Eq9}) yields the well-known
discrete quantization condition \cite{WKB1,WKB2,WKB3}
\begin{equation}\label{Eq14}
\int_{y_{t-}}^{y_{t+}}dy\sqrt{-V(y;M,Q,l,\mu,\alpha)}=\big(n+{1\over2}\big)\cdot\pi\
\ \ \ ; \ \ \ \ n=0,1,2,...\  .
\end{equation}
The two integration boundaries $\{y_{t-},y_{t+}\}$ of the WKB
formula (\ref{Eq14}) are the classical turning points [with
$V(y_{t-})=V(y_{t+})=0$] of the composed
charged-black-hole-massive-field binding potential (\ref{Eq11}) .
The resonant parameter $n$ (with $n\in\{0,1,2,...\}$) characterizes
the infinitely large discrete resonant spectrum
$\{\alpha_n(\mu,l,M,Q)\}_{n=0}^{n=\infty}$ of the black-hole-field
system.

Using the relation (\ref{Eq10}) between the radial coordinates $y$ and $r$, one can
express the WKB resonance equation (\ref{Eq14}) in the form
\begin{equation}\label{Eq15}
\int_{r_{t-}}^{r_{t+}}dr{{\sqrt{-V(r;M,Q,l,\mu,\alpha)}}\over{h(r)}}=\big(n+{1\over2}\big)\cdot\pi\
\ \ \ ; \ \ \ \ n=0,1,2,...\  ,
\end{equation}
where the two polynomial relations [see Eq. (\ref{Eq11})]
\begin{equation}\label{Eq16}
1-{{2M}\over{r_{t-}}}+{{Q^2}\over{r^2_{t-}}}=0\
\end{equation}
and
\begin{equation}\label{Eq17}
{{l(l+1)}\over{r^2_{t+}}}+{{2M}\over{r^3_{t+}}}-{{2Q^2}\over{r^4_{t+}}}
-{{\alpha Q^2}\over{r^4_{t+}}}=0\
\end{equation}
determine the radial turning points $\{r_{t-},r_{t+}\}$ of the composed black-hole-field binding potential
(\ref{Eq11}).

We shall now prove that the WKB resonance equation
(\ref{Eq15}) can be studied {\it analytically} in the regime (\ref{Eq13}) of large field masses.
To this end, it proves useful to define the dimensionless physical quantities
\begin{equation}\label{Eq18}
x\equiv {{r-r_{\text{+}}}\over{r_{\text{+}}}}\ \ \ \ ; \ \ \ \ \tau\equiv {{r_+-r_-}\over{r_+}}\  ,
\end{equation}
in terms of which the composed black-hole-massive-field interaction term (\ref{Eq11})
has the form of a binding potential well,
\begin{equation}\label{Eq19}
V[x(r)]=-\tau\Big({{\alpha Q^2}\over{r^4_+}}-\mu^2\Big)\cdot x +
\Big[{{\alpha Q^2(5r_+-6r_-)}\over{r^5_+}}-\mu^2\big(1-{{2r_-}\over{r_+}}\big)\Big]\cdot x^2+O(x^3)\  ,
\end{equation}
in the near-horizon region
\begin{equation}\label{Eq20}
x\ll\tau\  .
\end{equation}

From the near-horizon expression (\ref{Eq19}) of the
black-hole-field binding potential, one obtains the dimensionless
expressions
\begin{equation}\label{Eq21}
x_{t-}=0\
\end{equation}
and
\begin{equation}\label{Eq22}
x_{t+}=\tau\cdot{{{{\alpha Q^2}\over{r^4_+}}-\mu^2}\over{{{\alpha Q^2(5r_+-6r_-)}\over{r^5_+}}-\mu^2\big(1-{{2r_-}\over{r_+}}\big)}}\
\end{equation}
for the classical turning points of the WKB integral relation (\ref{Eq15}).

Taking cognizance of Eqs. (\ref{Eq20}) and (\ref{Eq22}), one finds that
our analysis is valid in the regime [see Eq. (\ref{Eq31}) below]
\begin{equation}\label{Eq23}
\alpha\simeq{{\mu^2r^4_+}\over{Q^2}}\  ,
\end{equation}
in which case the near-horizon binding potential and its outer
turning point can be approximated by the remarkably compact
expressions
\begin{equation}\label{Eq24}
V(x)=-\tau\Big[\Big({{\alpha Q^2}\over{r^4_+}}-\mu^2\Big)\cdot
x-4\mu^2\cdot x^2\Big]+O(x^3)\
\end{equation}
and
\begin{equation}\label{Eq25}
x_{t+}={1\over4}\Big({{\alpha Q^2}\over{\mu^2r^4_+}}-1\Big)\  .
\end{equation}
In addition, from Eqs. (\ref{Eq2}) and (\ref{Eq18}) one finds the near-horizon relation
\begin{equation}\label{Eq26}
h(x)=\tau\cdot x+(1-2\tau)\cdot x^2+O(x^3)\  .
\end{equation}

Substituting Eqs. (\ref{Eq18}), (\ref{Eq24}), (\ref{Eq25}), and
(\ref{Eq26}) into Eq. (\ref{Eq15}), one obtains the integral relation
\begin{equation}\label{Eq27}
{{1}\over{\sqrt{\tau}}}\int_{0}^{x_{t+}}dx \sqrt{{{{\alpha
Q^2}\over{r^2_+}}-\mu^2
r^2_+\over{x}}-4\mu^2r^2_+}=\big(n+{1\over2})\cdot\pi\ \ \ \ ; \ \ \
\ n=0,1,2,...\  .
\end{equation}
Defining the dimensionless radial coordinate
\begin{equation}\label{Eq28}
z\equiv {{x}\over{x_{t+}}}\  ,
\end{equation}
one can express the WKB resonance equation (\ref{Eq27}) in the mathematically
compact form
\begin{equation}\label{Eq29}
{{2\mu r_+ x_{t+}}\over{\sqrt{\tau}}}\int_{0}^{1}dz
\sqrt{{{1}\over{z}}-1}=\big(n+{1\over2})\cdot\pi\ \ \ \ ; \ \ \ \
n=0,1,2,...\  ,
\end{equation}
which yields the relation \cite{Noteintp}
\begin{equation}\label{Eq30}
{{\mu r_+ x_{t+}}\over{\sqrt{\tau}}}=n+{1\over2}\ \ \ \ ; \ \ \ \
n=0,1,2,...\  .
\end{equation}

From Eqs. (\ref{Eq25}) and (\ref{Eq30}), one finds the discrete
resonant spectrum
\begin{equation}\label{Eq31}
\alpha_n={{\mu^2r^4_+}\over{Q^2}}\Big[1+{{4\sqrt{\tau}}\over{\mu
r_+}}\big(n+{1\over2})\Big]\ \ \ \ ; \ \ \ \ n=0,1,2,...\
\end{equation}
for the dimensionless coupling parameter of the composed
charged-black-hole-non-minimally-coupled-linearized-massive-scalar-field
configurations in the regime $\mu r_+\gg1$ [see (\ref{Eq13})] of
large field masses. The analytically derived relation (\ref{Eq31})
can also be written as the discrete resonant formula \cite{Noteqr}
\begin{equation}\label{Eq32}
({{\mu
r_+}})_n=\sqrt{\alpha{{r_-}\over{r_+}}}-\sqrt{{{r_+-r_-}\over{r_+}}}\cdot\big(2n+1)\
\ \ \ \text{for}\ \ \ \ \alpha\gg1\
\end{equation}
for the dimensionless mass parameter which characterizes the
non-minimally coupled massive scalar field clouds in the
large-coupling $\alpha\gg1$ regime.

\section{Numerical confirmation}

It is of physical interest to test the accuracy of the analytically
derived resonant spectrum (\ref{Eq32}) in the large-coupling
(large-$\alpha$) regime against the corresponding exact ({\it numerically} computed \cite{aaa1,aaa2}) resonant spectrum.

In Table \ref{Table1} we present the {\it analytically} calculated
[see Eq. (\ref{Eq32})] dimensionless mass parameter $[\mu
r_+(\alpha)]^{\text{analytical}}$, which characterizes the composed
charged-black-hole-nonminimally-coupled-massive-scalar-field cloudy
configurations in the large-coupling $\alpha\gg1$ regime, for
various values of the dimensionless coupling parameter $\alpha$ of
the theory. We also present the corresponding exact (numerically
computed \cite{aaa1}) values of the dimensionless mass parameter
$[\mu r_+(\alpha)]^{\text{numerical}}$ which characterizes the
composed black-hole-field system. The data presented in Table
\ref{Table1} reveals the fact that the agreement between the
approximated (analytically derived) resonant spectrum (\ref{Eq32})
and the corresponding exact (numerically computed) resonant spectrum
becomes extremely good in the dimensionless large-coupling
$\alpha\gg1$ regime of the theory. In fact, the agreement between
the analytically derived resonant spectrum (\ref{Eq32}) and the
numerical results of \cite{aaa1} is found to be quite good already
in the $\mu r_+=O(1)$ regime.

\begin{table}[htbp]
\centering
\begin{tabular}{|c|c|c|c|c|c|c|}
\hline \ \ $\alpha(n)$\ \ & \ $82.52(0)$\ \ & \ $256.6(0)$\ \ & \
$469.8(0)$\ \ & \ $675.5(1)$\ \ & \ $2194(1)$\ \ & \ $4074(1)$\ \ \\
\hline \ \ $(\mu r_+)^{\text{analytical}}$\ \ &\ \
$2.80$\ \ \ &\ \ $5.63$\ \ \ &\ \ $7.94$\ \ \ &\ \ $7.88$\ \ \ &\ \ $16.39$\ \ \ &\ \ $23.33$\ \ \\
\hline \ \ $(\mu r_+)^{\text{numerical}}$\ \ &\ \
$2.75$\ \ \ &\ \ $5.60$\ \ \ &\ \ $7.92$\ \ \ &\ \ $7.88$\ \ \ &\ \ $16.39$\ \ \ &\ \ $23.33$\ \ \\
\hline
\end{tabular}
\caption{Composed
Reissner-Nordstr\"om-black-hole-non-minimally-coupled-massive-scalar-field
cloudy configurations. We display, for various values of the
dimensionless coupling parameter $\alpha$ of the theory and for
various values of the discrete resonant parameter $n$, the
analytically calculated values of the dimensionless mass parameter
$[\mu r_+(\alpha)]^{\text{analytical}}$ which characterizes the
composed black-hole-field system [see the analytically derived
resonant spectrum (\ref{Eq32})]. We also display the corresponding
exact (numerically computed \cite{aaa1}) values $[\mu
r_+(\alpha)]^{\text{numerical}}$ of the dimensionless mass parameter
of the composed black-hole-field system. The data presented is for
the case of a central supporting charged black hole with the
dimensionless charge-to-mass ratio $Q/M=0.7$. One finds that the
agreement between the {\it analytically} derived resonant spectrum
(\ref{Eq32}) and the corresponding {\it numerically} computed
spectrum \cite{aaa1} becomes extremely good in the dimensionless
large-coupling $\alpha\gg1$ regime of the composed black-hole-field
system. Interestingly, the data presented reveals the fact that the
agreement between the analytically derived resonant spectrum
(\ref{Eq32}) and the numerical results of \cite{aaa1} is quite good
already in the dimensionless $\mu r_+=O(1)$ regime.} \label{Table1}
\end{table}

\section{The $\alpha/\mu^2r^2_+\to r^2_+/Q^2$ limit}

Interestingly, it has been demonstrated numerically in
\cite{aaa1,aaa2} (see, in particular, Figure 4 of \cite{aaa1}) that
the dimensionless physical parameter $\alpha$ {\it diverges} in the
$\beta\to\beta_{\text{c}}$ limit, where the physical parameter
$\beta$ is defined by the dimensionless relation
\begin{equation}\label{Eq33}
{{\alpha}\over{\beta}}\equiv \mu^2 r^2_+\  .
\end{equation}
Here the critical parameter $\beta_{\text{c}}$ is given by the
simple relation \cite{aaa1}
\begin{equation}\label{Eq34}
\beta_{\text{c}}\equiv {{r^2_+}\over{Q^2}}\  .
\end{equation}

Remarkably, our results provide a simple {\it analytical}
explanation for the {\it numerically} observed \cite{aaa1,aaa2}
intriguing divergent functional behavior of the coupling parameter
$\alpha$ in the $\beta\to\beta_{\text{c}}$ limit. In particular,
from the resonant spectrum (\ref{Eq31}), one finds the simple
relation
\begin{equation}\label{Eq35}
\alpha_n={{16\beta_{\text{c}}\tau\big(n+{1\over2}\big)^2}\over{\big({{\beta}\over{\beta_{\text{c}}}}-1\big)^2}}\
\ \ \ \text{for}\ \ \ \ \beta\to\beta^{+}_{\text{c}}\  .
\end{equation}
We have therefore proved analytically that, in agreement with the
numerical results presented in \cite{aaa1,aaa2}, the physical
coupling parameter $\alpha$ of the composed black-hole-massive-field
system diverges quadratically in the $\beta\to\beta_{\text{c}}$
limit.

\section{Summary and Discussion}

The recently published highly important works
\cite{Herch,Herr,aaa1,aaa2} have explicitly proved, using numerical
techniques, that asymptotically flat charged black holes with
spatially regular horizons can support external static matter
configurations which are made of (massless as well as massive)
scalar fields. This physically intriguing phenomenon owes its
existence to a non-minimal coupling between the supported scalar
fields and the electromagnetic field of the charged black-hole
spacetime [see Eq. (\ref{Eq4})].

The interesting numerical results presented in
\cite{Herch,Herr,aaa1,aaa2} have revealed the fact that, in the
non-trivial field theory (\ref{Eq4}), the boundary between hairy
charged-black-hole-non-minimally-coupled-scalar-field configurations
and bald (scalar-less) Reissner-Nordstr\"om black holes is
determined by a critical existence-line $\alpha=\alpha_0(\mu;Q/M)$,
where the physical parameter $\alpha$ determines the strength of the
non-trivial coupling between the supported scalar configurations and
the electromagnetic field of the central charged black hole. In
particular, the critical existence-line of the system is composed of
{\it linearized} scalar field configurations which are supported by
central charged Reissner-Nordstr\"om black holes. Interestingly, it
has been demonstrated in \cite{Herch,Herr,aaa1,aaa2} that the
linearized external scalar configurations (scalar `clouds'), which
are supported in the charged black-hole spacetime, are characterized
by a discrete resonant spectrum
$\{\alpha_{n}(\mu;Q/M)\}_{n=0}^{n=\infty}$ of the non-trivial
scalar-field-electromagnetic-field coupling parameter $\alpha$.

In the present paper we have used {\it analytical} techniques in
order to explore the physical and mathematical properties of the
composed
Reissner-Nordstr\"om-black-hole-non-minimally-coupled-linearized-massive-scalar-field
system in the dimensionless large-coupling regime $\alpha\gg1$. In
particular, we have derived the discrete WKB resonant spectrum
(\ref{Eq32}) for the dimensionless mass parameter $\mu r_+$ which
characterizes the composed black-hole-massive-field cloudy
configurations. Furthermore, we have explicitly demonstrated that
the {\it analytically} derived resonant spectrum (\ref{Eq32}) for
the dimensionless mass parameter of the composed black-hole-field
system agrees remarkably well (see the data presented in Table
\ref{Table1}) with the corresponding numerically computed resonant
spectrum of \cite{aaa1}.

Interestingly, the analytically derived resonant formula (\ref{Eq32})
yields the remarkably compact expression \cite{Note00}
\begin{equation}\label{Eq36}
({{\mu r_+}})_{\text{max}}=\sqrt{\alpha{{r_-}\over{r_+}}}-
\sqrt{{{r_+-r_-}\over{r_+}}}\ \ \ \ \text{for}\ \ \ \ \alpha\gg1\
\end{equation}
for the critical existence-line which characterizes the
charged-black-hole-massive-scalar-field configurations.

The $\alpha$-dependent critical line (\ref{Eq36}) is physically
important in the composed Einstein-Maxwell-massive-scalar theory
(\ref{Eq4}) since it marks, in the large-mass regime, the boundary
between the hairy charged-black-hole-massive-scalar-field
configurations and the bald (scalar-less) Reissner-Nordstr\"om
black-hole spacetimes. In particular, for given parameters $\{M,Q\}$
of the central supporting black hole and for a given value of the
non-trivial coupling parameter $\alpha$, the hairy
charged-black-hole-massive-scalar-field configurations are
characterized by the critical inequality ${\mu}(\alpha;Q/M)\leq
{\mu}_{\text{max}}(\alpha;Q/M)$.

It is physically interesting to point out that the {\it
analytically} derived formula (\ref{Eq36}) for the critical
existence-line of the composed
charged-black-hole-massive-scalar-field system implies, in agreement
with the recently published important {\it numerical} results of
\cite{aaa1,aaa2}, that, for given values of the black-hole physical
parameters $\{M,Q\}$, the dimensionless mass parameter $\mu r_+$ of
the non-minimally coupled linearized scalar field configurations is
a monotonically increasing function of the dimensionless coupling
parameter $\alpha$ of the theory.

In addition, our analysis provides a simple {\it analytical}
explanation for the {\it numerically} observed \cite{aaa1,aaa2}
divergent functional behavior of the coupling parameter $\alpha$ in
the $\beta\to\beta_{\text{c}}\equiv r^2_+/Q^2$ limit, where
$\beta\equiv \alpha/\mu^2r^2_+$. In particular, using analytical
techniques, we have explicitly proved that the dimensionless
physical parameter $\alpha$ diverges {\it quadratically} [see Eq.
(\ref{Eq35})] in the $\beta\to\beta_{\text{c}}$ limit.

\bigskip
\noindent
{\bf ACKNOWLEDGMENTS}
\bigskip

This research is supported by the Carmel Science Foundation. I would
like to thank Yael Oren, Arbel M. Ongo, Ayelet B. Lata, and Alona B.
Tea for helpful discussions.



\begin{thebibliography}{99}
\bibitem{Noteqq} Here $Q$ is the electric charge of the charged black-hole spacetime.
We shall henceforth assume that $Q>0$ without loss of generality.
\bibitem{Noteunits} We shall use natural units in which $G=c=1$.
\bibitem{Notelm} For brevity, we shall henceforth omit the integer indices $\{l,m\}$
which characterize the spatially regular non-minimally coupled massive scalar field configurations.
\bibitem{Notemap} Note that the differential relation (\ref{Eq10}) maps the
regime $r\in[r_+,\infty]$ into the radial regime $y\in[-\infty,\infty]$.
\bibitem{Noteintp} Here we have used the integral relation $\int_{0}^{1}dx \sqrt{1/x-1}=\pi/2$.
\bibitem{Noteqr} Here we have used the identity $Q^2=r_+ r_-$ for the central supporting charged Reissner-Nordstr\"om black hole.
\end{thebibliography}
\end{document}